**UNIVERSITY OF OULU**

Center for Ubiquitous Computing

# Social Network Analysis of yahoo web-search engine query logs

Introduction to Social Network Analysis

18/04/2017

A H M Forhadul Islam

Mohamed Aboeleinen

# Contents

# List of Figures

# 1. Introduction

Web is now the undisputed warehouse for information. It can now provide most of the answers for modern problems. Search engines do a great job by combining and ranking the best results when the users try to search for any particular information. However, as we know "with great power comes great responsibility", it is not an easy task for data analysts to find the most relevant information for the queries. One major challenge is that web search engines face difficulties in recognizing users' specific search interests given his initial query.

In this project, we have tried to build query networks from web search engine query logs, with the nodes representing queries and the edges exhibiting the semantic relatedness between Queries.

Search Engines logs offer a very valuable source for all of the types of data. This data, which includes who did the search, when and where did he do it and much more than that, could be used to gain a lot insights and improve those search engines. However, it is very debatable whether it is ethical to use this type of data because of privacy concerns.

Our proposed model uses only the search queries without linking them to any type of data linked to the users and tries to detect patterns and trends which could be useful for both understanding the behavior of users and improving the efficiency of search engines.

# 2. Glossary

Semantic Relatedness: could be described as the semantic relationship between units of language, concepts or instances.

Betweenness Centrality: In graph theory, betweenness centrality is a measure of centrality in a graph based on shortest paths.

Degree of the Node: It is the number of edges incident on a certain node with loops counted twice.

NLTK: Natural Language Toolkit. NLTK is a platform for building Python programs to work with human language data. It provides interfaces for WordNet, along with a suite of text processing libraries for functions as classification, tokenization, stemming and tagging.

## 3.  State of art

A number of paper have been published on finding semantic relatedness between the search engine query logs.

Xiaodong Shi (Shi, X., 2007), proposed to segment users[1] query histories into sessions of queries and calculate semantic relatedness using some statistical measures including collocation, weighted dependence and mutual information, of queries co-occurring in these query sessions. He also tried to develop an undirected network of queries based on the modeled semantic relatedness between queries. Such a query network consists of queries as nodes and the semantic relatedness between queries as edges. After that he examined the properties of constructed query networks and concluded that the query networks are typical small world networks.

Fonseca, et al. (Fonseca et al., 2003) proposed a data mining approach with extracting association rules of queries from query logs. They divided users' query histories into query sessions from which relation rules were obtained. Relevancies of the queries were ranked after that. Later in another work, Fonseca, et al. (Fonseca et al., 2005) proposed to build simple query graphs based on association rules and to detect cliques in query graphs. Identified cliques are associated with query concepts.

Another interesting approach by Hecht, et al. (Hecht, et al., 2012)[9] introduces helpful ideas in the visualizing relatedness between queries. They built a model that generates interactive visualizations of query concepts using thematic cartography (e.g. choropleth maps, heat maps).Their results could be linked to geographical maps, illustrations as periodic tables. This works on a predefined concept, for instance elements that is used in nuclear energy and then links it to a predefined illustration, in this case it would be the periodic table.

Benz, et al. (Benz et al., 2009) has done a comparison between the semantic relatedness deduced from folksonomies, in this case explicit tagging in social bookmarking, on one hand and logsonomies (logs of the clicks made by users when using search engines) and resource context relatedness on the other hand and how they could be used together for better optimization of search engine results. Folksonomy alongside logsonomy was proven to give better results.

Most of the research approaches to make use of query logs relies heavily on the use of users' data such as clicks, location, demographic data or data about their search sessions. This can cause some privacy issues and that is why the analysis of query logs alone could be an interesting point that needs investigation.

# 4. Data Set

The dataset was published by Yahoo under the name L13 - Yahoo! Search Query Tiny Sample[8]. According to the description they provided: This dataset contains a random sample of 4496 queries posted to Yahoo's US search engine in January, 2009. For privacy reasons, the query set contains only queries that have been asked by at least three different users and contain only letters of the English alphabet, sequences of numbers not longer than four numbers and punctuation characters. The query set does not contain user information nor does it preserve temporal aspects of the query log. Total size for this dataset is 41K.

This dataset contains a random sample of 4496 queries posted to Yahoo's US search engine in January, 2009. The query set contains only queries that have been asked by at least three different users and contain only letters of the English alphabet, sequences of numbers not longer than four numbers and punctuation characters.

# 5. Methodology

The steps that we took in order to analyze the dataset was mainly: data preparation, data filtering, evaluating semantic analysis and finally visualization. Here is a simple pseudo code that describes the whole operation.

01.    procedure Semantic relatedness of Yahoo query logs
02.    Import data
03.    Split Queries
04.    Tokenize Query Word

| | | |
|---|---|---|
| 05. | For each query pair | |
| 06. | Detect parts of speech | |
| 07. | for each noun pair | |
| 08. | | evaluate semantic relatedness of the pair using word net |
| 09. | | evaluate edit distance |
| 010. | for each verb pair | |
| 011. | | evaluate semantic relatedness of the pair using word net |
| 012. | | evaluate edit distance |
| 013. | Calculate relatedness between this query pair | |

## Dependencies

- Python 2.7
- Nltk library
- python-arango client

## Database

- ArangoDB
- SQlite

ArangoDB was used for initial graph representation. SQLite was used for storage of nodes and edges.

## Algorithm Steps

We have created our own method for finding semantic similarities between two strings. The method named **semantic_similarity()** is written for that purpose. It calculates the noun weights, verb weights and edit distance value of the two strings. After that these values are added and the total weight is returned. The detailed steps are as following:

1. We started by tokenizing the queries as some of the queries had more than one word.
2. After tokenizing we used pos_tag to see what part of speech each word is.

3. Then we compared similar parts of speech for example: nouns with nouns and verbs with verbs to see their semantic relatedness using wordnet. The scores are added and then normalized.
4. We also calculated the edit distance and if it is less than a certain score it is added to the total score.
5. If the total score is above a certain threshold then an edge between this pair of queries.
6. This was repeated for each set of pairs of queries.
7. Unusual keywords ( urls, numbers ) were removed from the dataset.

After filtering the keywords, all of the possible pairs of keywords were evaluated. The relatedness of each possible pair of keywords were evaluated and given a weight. If the weight is above a certain threshold, an edge is created between those nodes.

## Code

```python
def semantic_similarity(str1, str2):
    tk1 = nltk.word_tokenize(str1)
    pt1 = nltk.pos_tag(tk1)
    tk2 = nltk.word_tokenize(str2)
    pt2 = nltk.pos_tag(tk2)

    isUrl1 = is_valid_url(str1)
    isUrl2 = is_valid_url(str2)
    if isUrl1 or isUrl2:
        return 0

    nounWeight = 0
    verbWeight = 0
    outputWeight = 0
    s = 0

    cd1 = [word for word,pos in pt1 if pos == 'CD']
    cd2 = [word for word,pos in pt2 if pos == 'CD']

    if cd1 or cd2:
        return 0

    # Calculating noun weights
    propernouns1 = [word for word,pos in pt1 if pos == 'NN' or pos == 'NNP' or pos == 'PRP']
```

```python
    propernouns2 = [word for word,pos in pt2 if pos == 'NN' or pos == 'NNP' or pos ==
    'PRP']

    if propernouns1 and propernouns2:
        for pn1 in propernouns1:
            syns1 = wn.synsets(pn1)
            for pn2 in propernouns2:
                d = nltk.edit_distance(pn1, pn2)
                syns2 = wn.synsets(pn2)
                try:
                    s = syns1[0].wup_similarity(syns2[0])
                    if s:
                        s = s / (len(propernouns1) + len(propernouns2))
                        nounWeight += s

                    # If distance is less than or equal 2 then add 0.2
                    if d <= 2:
                        nounWeight += 0.2
                except:
                    continue

    # Calculating verb weights
    verbs1 = [word for word,pos in pt1 if pos == 'VBN' or pos == 'VB']
    verbs2 = [word for word,pos in pt2 if pos == 'VBN' or pos == 'VB']
    s = 0
    if verbs1 and verbs2:
        for vb1 in verbs1:
            syns1 = wn.synsets(vb1)
            for vb2 in verbs2:
                d = nltk.edit_distance(vb1, vb2)
                syns2 = wn.synsets(vb2)
                try:
                    s = syns1[0].wup_similarity(syns2[0])
                    if s:
                        s = s / (len(verbs1) + len(verbs2))
                        verbWeight += s

                    # If distance is lesst than or equal 2 then add 0.2
                    if d <= 2:
                        verbWeight += 0.2
                except:
                    continue

    outputWeight = nounWeight + verbWeight
    return outputWeight
```

We have also removed all the urls from the keywords as it would be difficult to fetch the information about the domains and comparing the information with other keywords. Here is the method is_valid_url() which was used for detecting url.

```python
def is_valid_url(url):
    import re
    u = url.split(' ')
    if len(u) > 1:
        return None
    regex = re.compile(
        r'((?:[A-Z0-9](?:[A-Z0-9-]{0,61}[A-Z0-9])?\.)+[A-Z]{2,6}\.?|)'
        r'(?::\d+)?'
        r'(?:/?|[/?]\S+)$', re.IGNORECASE)
    return url is not None and regex.search(url)
```

To find semantic similarity between two strings we can then call the semantic_similarity method and passing two strings for which we need to calculate the semantic similarity. Here is a sample code for calling the method:

```
print semantic_similarity('world war','the great war')
```

When this line of code is written in the console then we will get a value which is the total weight of similarity between the two strings.

## Visualization

Initially we have tried to visualize the graph using ArangoDB which is a graph database but there are some limitations with the database as we needed to calculate various factors regarding the relationship between the nodes. So, later we have used Gephi for the visualization. Gephi provides a lot of functionalities for calculating the factors between nodes.

NodeXL was used for extracting the adjacency matrix and then used UCInet to calculate the betweenness centrality of the nodes. After that we used the degree centrality to visualize clusters "sub graphs" using NodeXL.

# 6.  Results and Discussions

As was mentioned before, the aim of the project was to analyze the semantic relatedness of search queries without taking into account any of the users' information e.g. location, age, time and create some visualization that would make it easier for analysts to find trends that would help in search engines optimization.

We used Gephi at first, "Gephi is a visualization and exploration software for of graphs and networks", but it was hard to extract insights from the visualization. So we had to switch from Gephi to NodeXL.

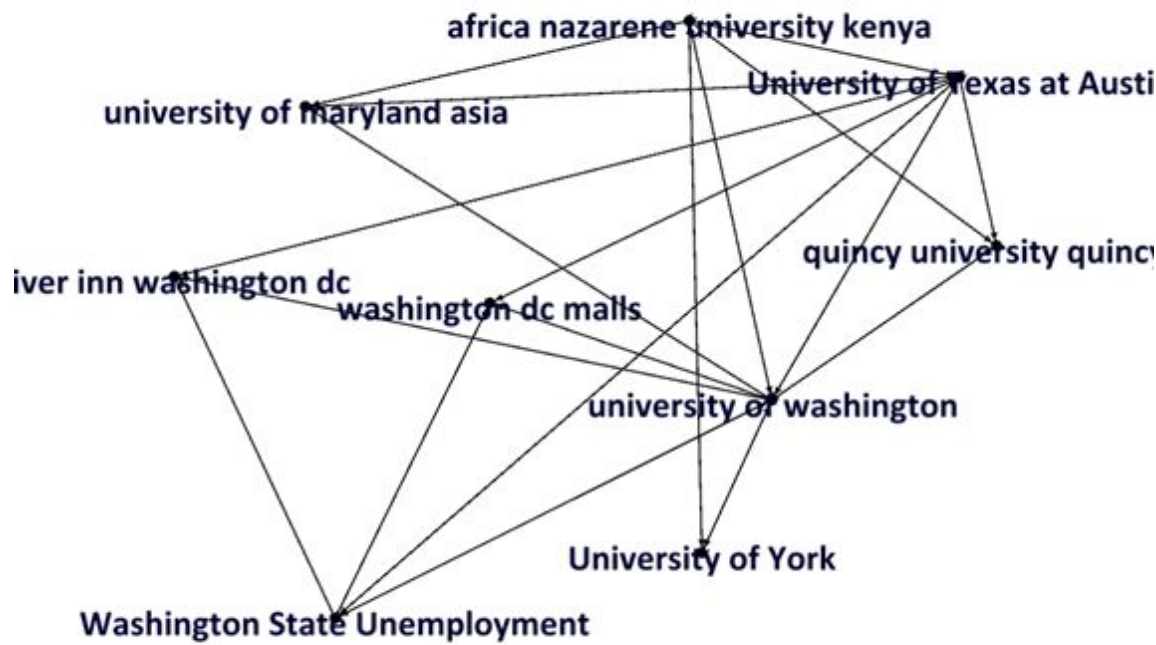

Figure 1: Graph of Semantically related queries to University of Washington visualized by Gephi

Figure 1, shows the semantically related queries to University of Washington. It is noticeable that some of the queries is related to Washington as a place, for example: iver inn washington dc, washington dc malls, or related to universities, for example: University of York, university of Maryland asia. However the method failed to detect the difference between Washington D.C and State of Washington. University of Washington is not in Washington D.C. Accordingly, it would have made more sense if it was connected to places in Washington State and not the places in Washington D.C.

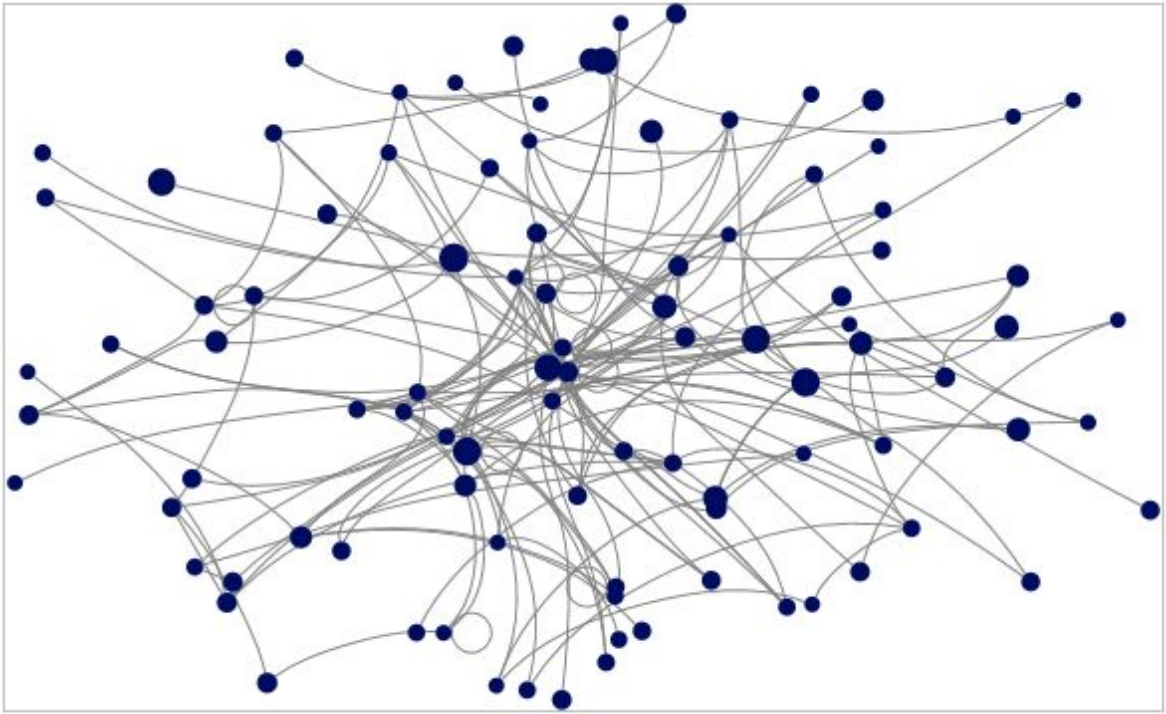

Figure 2: Graph of Semantically related queries before applying clustering

After switching to NodeXL, we used the degree centrality to visualize clusters "sub graphs". In order to gain better insights we had to show various types of information on the same graph. This enables analysts to get better insights. The information that was shown includes:

·       **Betweenness centrality**:  the color of the node represents betweenness centrality, i.e. red is high, green is low.

·      **Degree of the node**: the degree of the node determines whether the node will be shown in the graph or not, i.e. the higher the degree, the higher the possibility it would show in the graph also the size of the node shows its degree, i.e. the higher the degree, the bigger the node will be.

·       **Total weight** "calculated by our algorithm": this is represented in the width of the edge, i.e. the thicker the edge the higher the weight.

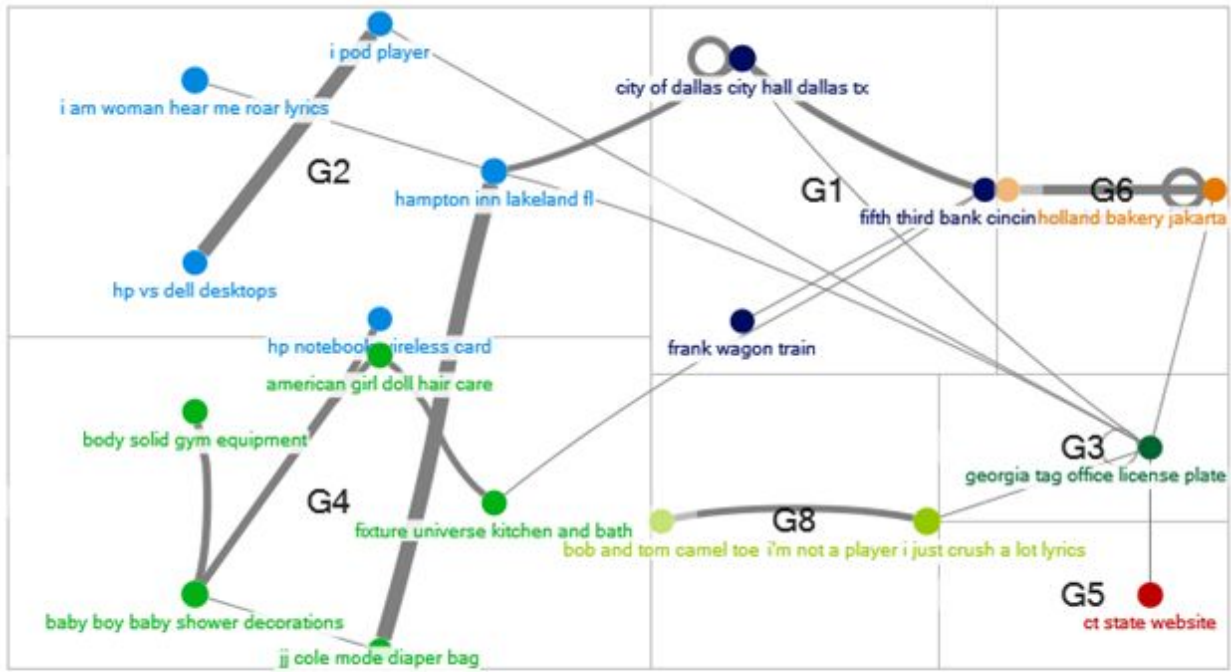

Figure 3: Graph showing semantically related queries clustered in groups with degree of the node, total weight and betweenness centrality visualized.

For example, figure 2 shows a semantically related subset of the queries. However it is really hard to get insights from this kind of graph. On the other hand, when we apply clustering and visualize more information on the graph (degrees of the node, betweenness centrality, total weight) as in figure 3, some properties of the graph starts to appear.

In figure 1, we decreased the number of nodes by visualizing nodes with the highest betweenness centrality so we can analyze it in an easy way. Then as we go through our analysis, the number of nodes will increase. Figure 3 shows a number of groups of semantically related queries. For instance we can observe that G1 shows queries related to places in U.S.A (city of **dallas** city hall **dallas** tx, fifth third bank **cincin**[1], frank wagon train[2]). G2 showing queries that could be considered related to electronics and music (**hp** vs dell **desktops**, i am a woman hear me roar **lyrics**, **ipod** player) and hampton inn lakeland which is not clear to us how is it related to this group. It is also visible how the two queries (hp vs dell desktops) are highly related to (i pod player) which can be seen in the width of the edge connecting the two nodes.

---

[1] Fifth Third Bank is a U.S. regional banking corporation, headquartered in Cincinnati, Ohio at Fifth Third Center.
[2] Adventures of a wagon train traveling between two places in U.S.A Missouri to California.

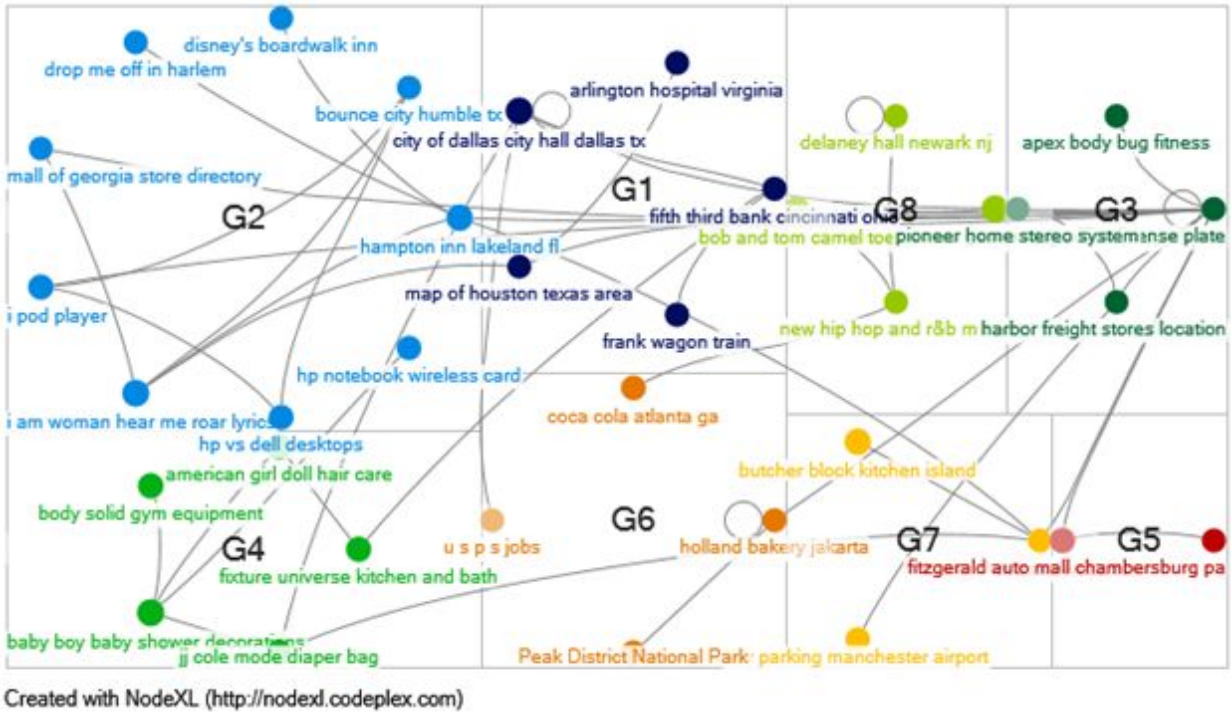

Figure 4: The Graph after increasing the number of nodes visualized.

As the graphs started to make better sense to us, we started increasing the number of nodes (queries) displayed in the graph. As shown in figure 4, we increased the number of nodes and we can observe that some of the nodes were added to new groups (cluster) and the others were added to the existing groups. If we took G2 as an example, we can see that more nodes were added following the seemingly same behaviour in figure 3. For instance nodes as (hp notebook wireless card) have been added to G2. "Map of Houston" was added to G1. Queries in G4 (baby boy baby shower decorations, future kitchen and bath, american girl doll hair care) can be considered related to housekeeping and babies

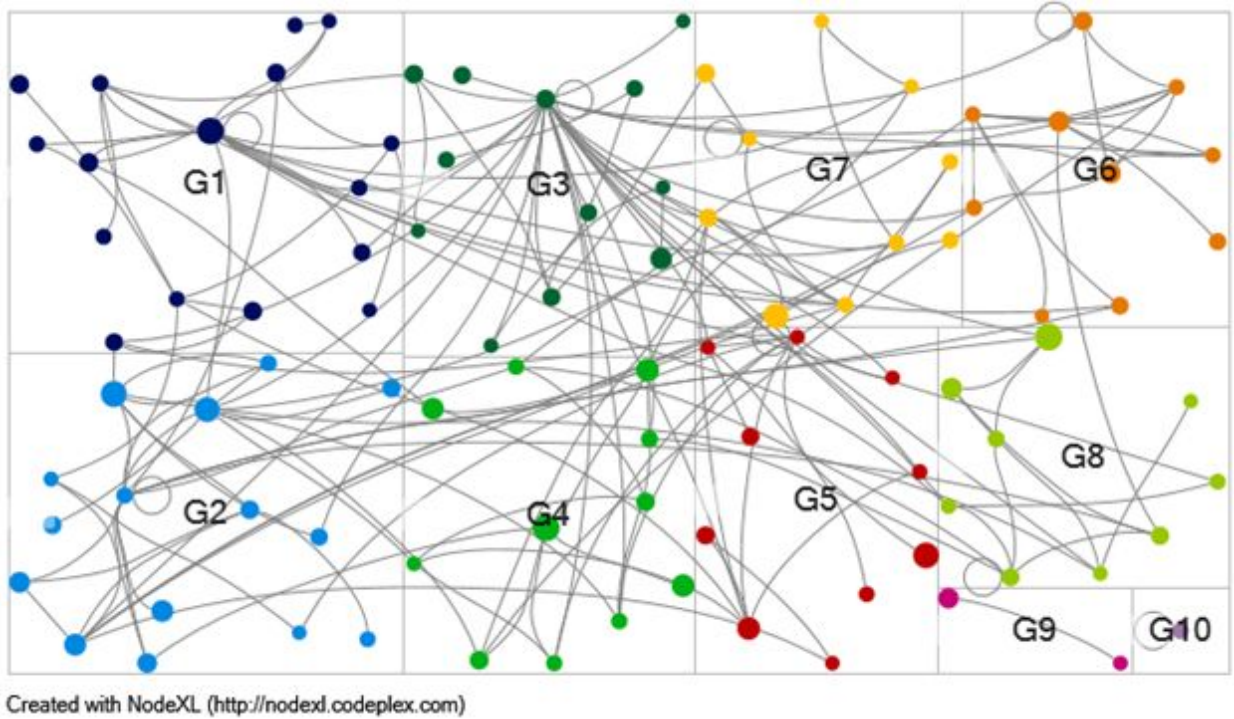

Figure 5: A graph with more nodes visualized by the same method

# 7.  Conclusion

In this project, we worked on the analysis of small dataset having only the search queries of yahoo search engine. We examined the possibility of getting useful insights by analysis the semantic relatedness among those queries only. Most of the models that are built to analyze search queries uses users' data in their analysis which causes concerns about privacy issues.

For this purpose we built our model which examines the relatedness between the search queries based on the edit distance between similar parts of speech and on the wordnet semantic relatedness. Afterwards, we created a graph where each node is search query and each edge resembles the relatedness that we calculated. The graph was visualized and clustered in order to get some useful insights from the data.

Our model was able to detect similarities between search queries and group them in related groups. These groups could be labeled or categorized and thus would give analysts a method to summarize and categorize search queries without needing to access users' data as proposed by Xiaodong Shi (Shi, X., 2007)[1].

# 8. Future Work

Our model was able to evaluate the relatedness among search queries and then group them is semantically related clusters. This could be further utilized by using a libraries as SpaCy[3] to name or label those categories. It would be possible using this library we can put label on the keywords.

Another method that could increase the accuracy of our semantic relation analysis method is to consider the search engine results to see if the results that is related to the different queries are similar or not. This could help in finding the relations between queries that does not have similarities in their roots or are not considered similar using wordnet.

We can calculate Jaccard index (also known as Jaccard similarity coefficient) which compares two sets to see if they are similar and distinct. It's a measure of similarity for the two different set of data. The value has a range of 0% to 100%. The higher the percentage, the more similar the two keywords. For our case, we can easily find similar keywords using Jaccard index.

---

[3] SpaCy library: [ https://spacy.io ]